# Stanene: A Good Platform for Topological Insulator and Topological Superconductor

Chenxiao Zhao [a], Jinfeng Jia [*a,b,c]

[a] *Key Laboratory of Artificial Structures and Quantum Control (Ministry of Education), Shenyang National Laboratory for Materials Science, School of Physics and Astronomy, Shanghai Jiao Tong University, Shanghai 200240, China*

[b] *Tsung-Dao Lee Institute, Shanghai Jiao Tong University, Shanghai 200240, China*

[c] *CAS Center for Excellence in Topological Quantum Computation, University of Chinese Academy of Sciences, Beijing 100190, China*

*Corresponding author. E-mail: jfjia@sjtu.edu.cn*

**Abstract:**

Two dimensional (2D) topological insulators (TIs) and topological superconductors (TSCs) have been intensively studied for recent years due to its great potential for dissipationless electron transportation and fault-tolerant quantum computing, respectively. Here we focus on stanene, the tin analogue of graphene, to give a brief review of its development as a candidate for both 2D TI and TSC. Stanene is proposed to be a TI with a large gap of 0.3 eV, and its topological properties are sensitive to various factors, e.g., the lattice constants, chemical functionalization and layer thickness, which offer various methods for phase tunning. Experimentally, the inverted gap and edge states are observed recently, which are strong evidence for TI. In addition, stanene is also predicted to be a time reversal invariant TSC by breaking inversion symmetry, supporting helical Majorana edge modes. The layer-dependent superconductivity of stanene is recently confirmed by both transport and scanning tunneling microscopy measurements. This review gives a detailed introduction to stanene and its topological properties and some prospects are also discussed.

Contents

# 1  Introduction

Topological insulator (TI) and topological superconductor (TSC) are ideal systems for electronic transport and topological quantum computing, respectively. The two dimensional (2D) version of the TI is also called quantum spin hall (QSH) insulator, which holds metallic edge states[1-5]. These states are immune to backscattering, thus tremendously reducing the energy loss. Stanene, whose structure is the same with monolayer α-Sn, was first proposed to be a large gap QSH insulator in 2013 by Zhang's group[6]. The inverted gap of 0.3 eV is large enough to sustain room-temperature QSH effect, which is essential for practical topological transport devices. After that, numerous works are reported to explore its topological properties. Theoretically, it is found that its band structure and topological properties are tunable by various factors, e.g., substrates[7-9], chemical functionalization[6, 10] and film thickness[11], which provides numerous approaches for phase tuning. The experimental confirmations are far behind theory, which is probably because of its less-stable structure and strain-sensitive topological properties. Some experimental results of stanene show a trivial insulating gap[12, 13], and some even acquire metallic band structure[14, 15]. The nontrivial band gap of stanene is observed until 2018, when growing stanene on Cu (111)[16]. Recently, the spatial distribution of edges states (STM evidences) have been observed, providing new evidence on its QSH property[17, 18].

2D TSC is characterized by triplet-pairing superconducting gap in the bulk, and self-conjugated excitations called Majorana mode at their boundaries, which can be used for topological quantum computing[4, 5, 19-21]. Zhang *et al.* proposed that Ag-doped stanene can be a time-reversal-invariant TSC, where helical Majorana modes reside at edges [22, 23]. However, previous experiments show that the bulk α-Sn is not a superconductor. Fortunately, when reducing to 2D (few layers stanene), it shows remarkable superconductivity which is reported recently by transport measurements and STM studies, but its odd parity is still unconfirmed [18, 24].

This review gives an overview of the theoretical calculations and experiments for topological and superconducting properties of stanene. After this introduction, the structure of stanene is introduced in Section 2. Then its topological properties to be a 2D TI and 2D TSC are discussed in Section 3 and 4 respectively. After that, some other topological properties of stanene, such as to become a quantum anomalous hall insulator and 3D Dirac semimetal, are also mentioned. Finally, we give a summary and perspective for stanene system.

## 2 Structure of stanene

This section discusses the atomic structure of stanene in QSH phase as well as on different supporting substrates.

### 2.1 From graphene to stanene

The birth of graphene ignited researchers' enthusiasm for 2D materials[25]. Since that, theoretical calculations and experimental measurements on graphene are booming in this field. Figure 1(a) gives a visualized structure of graphene, where the carbon atoms arranged as a honeycomb lattice. From the side views we can see that all carbon atoms lie in the same plane without buckling. This plane geometry is attributed to the strong π-π bonding force, far exceeding the thermal fluctuations, between carbon atoms, which makes the plane structure avert falling out.

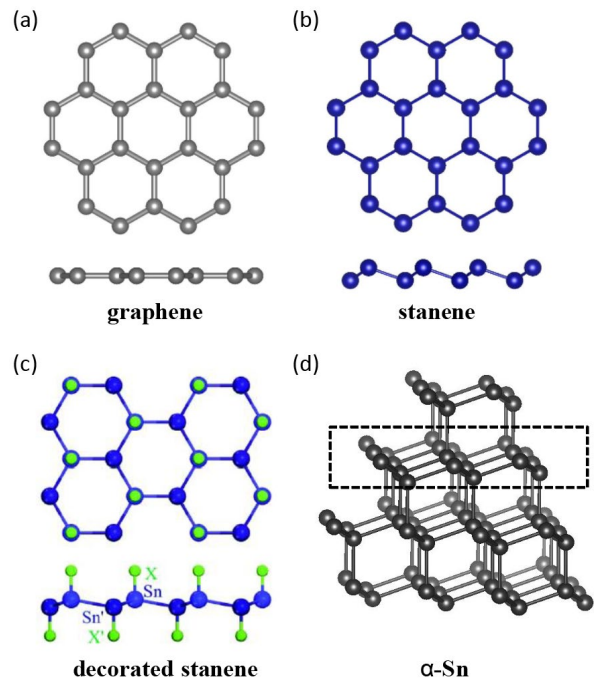

FIG. 1. Structure of graphene and stanene. (a) Top view (upper panel) and side view (lower panel) of graphene. (b) Top view (upper panel) and side view (lower panel) of stanene. (c) Top view (upper panel) and side view (lower panel) of decorated stanene. Reproduced from

Ref [6]. (d) Bulk structure of α-Sn. Single layer of α-Sn (111) is marked out by black dotted rectangle.

As an analogue of graphene, stanene also has a honeycomb structure, but in a buckling form [see Fig. 1(b)]. The reason for this buckling is that the relatively weak π-π bonding, resulting from large bond length between tin atoms, can not stabilize the planar configuration. The overlap between π orbital and σ orbital is enhanced via the buckling, and thus stabilizes the system[6]. Due to the buckling, stanene has two sub-lattices of tin atoms, shifting along $Z$ direction [see lower panel of Fig. 1(b)]. The lattice constant calculated by Density Functional Theory (DFT) is 0.467nm. It should be noted that the stanene has dangling bonds on both sides [see Fig. 1 (c)], then it is inclined to have chemical passivation, forming a more stable $sp_3$ configuration. In fact, stanene is not a newly-established structure of tin atoms. It can be regarded as single layer α-Sn (111) [see Fig. 1(d)], which is a pre-existing phase of Sn.

### 2.2 Lattice structure obtained in experiments

Although theoretical calculations confirm the stability of stanene, it is difficult to achieve high-quality stanene films in experiments. Molecular Beam Epitaxy (MBE) is the most suitable method to grow such thin films. The first hurdle to overcome is to find a suitable substrate for MBE growth, which provides small lattice mismatch. An improper substrate lattice would highly likely lead to the formation of β-phase Sn or even amorphous cluster of Sn. The first atomically flat stanene films are successful grown on Bismuth telluride (111) [$Bi_2Te_3$ (111)] surface [see Fig. 2(a)] [14]. $Bi_2Te_3$ (111) surface has hexagonal lattice with lattice constant of 0.435 nm, slightly smaller than that of stanene. The stanene grown on top has the same lattice constant with substrate, indicating a compressive strain is applied. This mismatch leads to a highly buckled stanene surface. The atomically resolved image by scanning tunneling microscopy (STM) gives a clear sight of upper tin atoms [see Fig. 2(b)]. After that, other substrates with similar lattice constant with $Bi_2Te_3$ are used to grown stanene, such as PbTe (111)[12], Sb (111) [26]and InSb (111) [13]. Notably, all stanene films in these works suffer a compressive strain and have a highly buckled configuration. As a result, only the top sub-layer tin atoms can be distinguished.

Considering the stanene and the substrate as a combined system, an effective way to get more flat stanene is to find a substrate with which the combined system has lowest energy when stanene is in flat configuration. Some metal surfaces, e.g., Ag (111) and Cu (111), meet this requirement[16, 27, 28]. Recently, ultra flat stanene films are grown on Cu (111) surface [see Fig. 2(e)], and the honeycomb structure is clearly observed by STM [see Fig. 2(f)] [16]. The lattice constant measured is 0.51 nm, much larger than the theoretical value. This large lattice leads to a nearly plane structure.

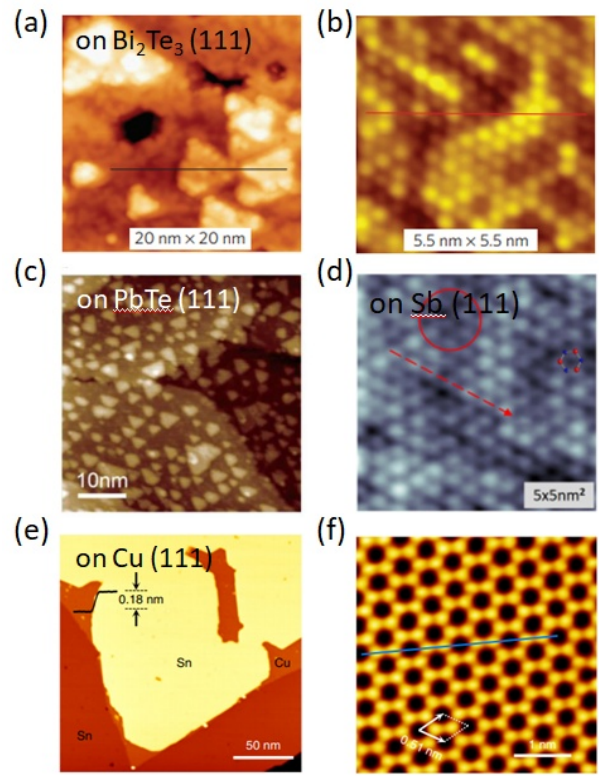

FIG. 2. Stanene films grown on different substrates. (a) Topography of stanene films grown on $Bi_2Te_3$ (111) by STM. Reproduced from Ref [14]. (b) Atomically resolved image of top sub-layer tin atoms of stanene grown on $Bi_2Te_3$ (111). Reproduced from Ref [14]. (c) Topography of stanene films on PbTe (111) by STM. Reproduced from Ref [12]. (d) Atomically resolved image of top sub-layer tin atoms of stanene grown on Sb (111). Reproduced from Ref [26]. (e) Topography of stanene films on Cu (111) substrate by STM. Reproduced from Ref [16]. (f) Atomically resolved image of stanene grown on Cu (111). Reproduced from Ref [16].

At the end of this part, we summarize the measured lattice values of stanene on different substrates in Table 1, with the α-Sn (111) surface and freestanding stanene as references (marked in red). It is easily to find that stanene always keeps the same lattice constant with their substrate. Thus the stanene grown on primitive (1×1) substrates usually has a compressed lattice. While on reconstructed substrate surface, stanene can have a large lattice, leading to an ultra flat honeycomb configuration. This provides us a guidance for controlling lattice constants via substrate.

Table 1. lattice constants of stanene grown on different substrates

| substrate | Lattice of substrate (nm) | Lattice of stanene (nm) |
|---|---|---|
| Sb (111) [26] | 0.43 | 0.43 |
| $Bi_2Te_3$ (111) [14] | 0.435 | 0.435 |
| PbTe (111) [12] | 0.452 | 0.452 |
| Bi (111) [29] | 0.454 | 0.45 |
| InSb (111) [13] | 0.458 | 0.458 |
| α-Sn (111) | None | 0.46 |
| free standing stanene [6] | None | 0.467 |
| Ag (111) $\sqrt{3} \times \sqrt{3}$ [28] | 0.509 | 0.509 |
| Cu (111) $2 \times 2$ [16] | 0.51 | 0.51 |

## 3  Stanene as topological insulator

This section focuses on the topological properties of stanene to be a 2D TI. A brief introduction for 2D TI is given at first; then the theoretical predictions for stanene to be a 2D TI are introduced; finally, the existing experimental results are described. What is more, some factors are discussed to tune its electronic and topological properties.

### 3.1  Development of 2D topological insulator

2D TIs, also called QSH insulators, are characterized by insulating gap in the bulk and conducting channels without backscattering at edges. And each edge contributes a quantized conductance of 2e²/h[4, 5, 30]. This dissipationless conducting channels can tremendously reduce the power dissipation and have great potential for low-consumption electronic devices. The first predicted 2D topological insulator is graphene. At low temperature, spin orbital coupling (SOC) induced energy gap converts graphene from a semimetal to an insulator, and helical edge states reside at boundaries[2]. However, the SOC in graphene is extremely weak, giving an insulating gap as tiny as $10^{-3}$ meV[31]. This small gap brings difficulties and challenges for experimental observation and practical applications. Since that, a lot of materials with larger nontrivial gaps are proposed to realize quantum spin hall effect.

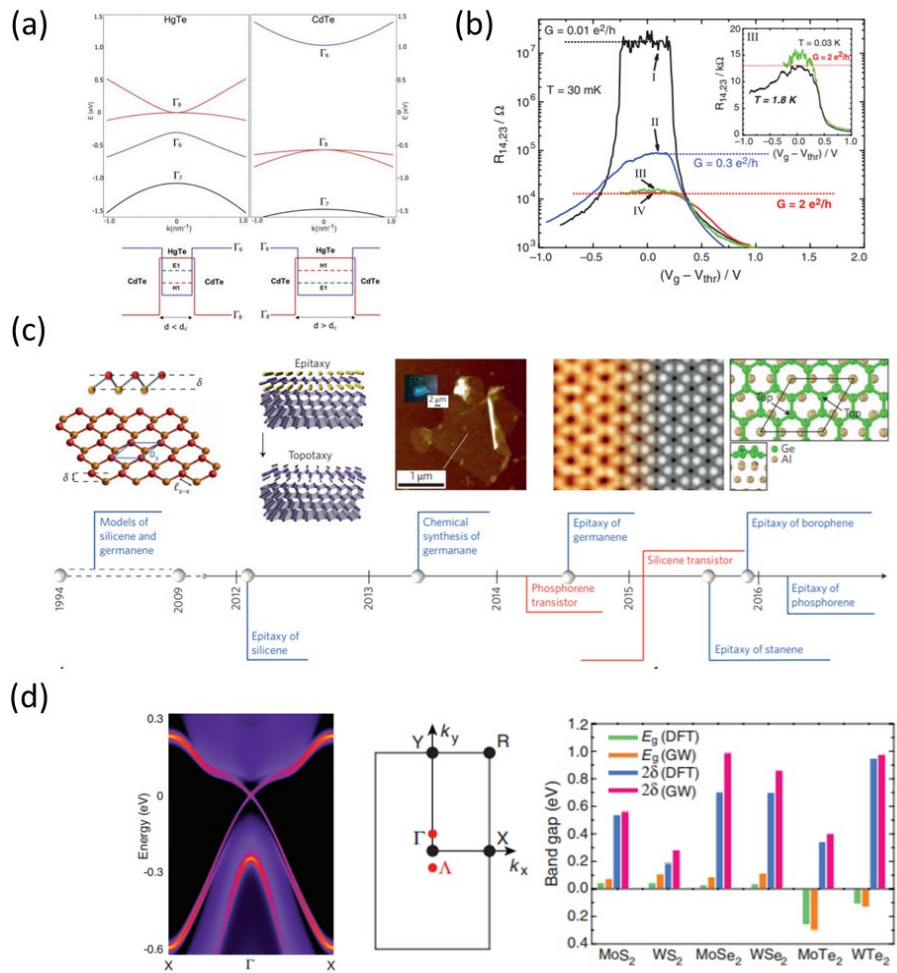

FIG. 3. Candidates for 2D TI. (a) Upper pannel: Band structures of HgTe and CdTe. Lower pannel: Width-dependent band structures of HgTe/CdTe quantum well. Reproduced from Ref [32]. (b) Experimental observation of the quantized edge conducting channel in HgTe/CdTe quantum well. Reproduced from Ref [3]. (c) The development of Xene. Reproduced from Ref

[33]. (d) Theoretical prediction of the inverted band gaps for monolayer 1T'-TMDs. Reproduced from Ref [34].

The first successful system is HgTe/GdTe quantum well [see structure in Fig. 3 (a)][32]. HgTe has an inverted band structure with the p type $\Gamma_8$ band lying above the s type $\Gamma_6$ band, which is shown in figure 3 (a). This band inversion in HgTe is due to the spin-orbit splitting of the p orbitals as well as the relativistic mass-velocity correction of the 6s electrons of Hg[35]. On the contrary, CdTe has the normal band structure that $\Gamma_8$ band lies below the $\Gamma_6$ band. When the thickness of HgTe surpassing a critical value $d_c$, the quantum well possesses an inverted band structure, i.e., in the quantum spin hall phase. The transport evidences were reported in the following year [see Fig. 3(b)], confirming the prediction[3]. However, this sandwich structure is hard to fabricate and the conductance plateau exists only at ultra-low temperature (about 30 mK), hindering practical applications of HgTe/GdTe quantum well at room temperature. As a result, finding alternatives to realize high temperature quantum spin hall effect is one of the main tasks in this field.

Plenty of candidates are proposed in the next few years, such as graphene-like Xene [33], monolayer 1T'-phase transition metal dichalcogenides (TMDs)[34], layered transition metal pentatelluride $XTe_5$[36], etc. Among these candidates, Xene family and 1T'-TMD family are the most promising materials. Xene family is an extension of graphene, which use heavy elements to substitute carbon[see figure 3(c)] to enhance SOC. Owing to the strong SOC, Xene family generally has a large insulating gap of several hundreds meV[6, 37], which is sufficient large for practical applications. But the buckled honeycomb structure is less-stable and difficult to grow. On the contrary, 1T'-phase TMD materials are more stable once successfully grown. But the insulating gaps of 1T'-TMD family are commonly less than 100 meV[38, 39], hindering its application prospects. Recently, the topological properties of the 1T'-WTe$_2$ is confirmed by several experiments, including STM[40, 41], angle-resolved photoemission (ARPES)[38] and transport measurements [42, 43]. While for the highly expected stanene, different experiments show distinct results and its topological properties are still under debate. Despite that, stanene is a good platform for 2D TI for its large insulating gap and simple structure with just single element.

### 3.2 Theoretical prediction for stanene to be a 2D topological insulator

Stanene was first proposed to be a 2D TI in 2013 by Shoucheng Zhang's group[6]. At first, only freestanding condition and chemical functionalizations were taken into consideration. Fig. 4(a) shows the band structure of bare stanene without decoration, where a Dirac point locates at K point without SOC [see Fig. 4(a)]. After involving SOC effect, an insulating gap opens with inverted band structure, rendering bare stanene to be a 2D TI with a gap ~ 0.1 eV[6]. In the bare case the mechanism for band inversion is the same with that in graphene, and the dramatically enhanced SOC effect results from both the huge contribution from first order SOC which is banned in graphene, and the heavier Sn atoms[44]. What is more, the dangling bonds of stanene provide an additional degree of freedom to tune the band structure including topological properties. When decorated by chemical functional group, the $p_z$ orbital is saturated and the dominated states by it at K point will open a large gap of several electron volts [see figure 4 (b)]. Thus only the band at Γ point should be considered. Before chemical decoration, the parities of occupied states at Γ point are unchanged, which denotes a topological trivial phase. An s-p band inversion, similar with that in HgTe[45, 46] and bulk α-Sn cases[47, 48], happens after the chemical decoration, giving a nontrivial band structure. Finally, the SOC effect opens a gap, driving decorated stanene into QSH phase with a large gap of ~ 0.3 eV.

Take fluorinated stanene as an example, the band topology of decorated stanene is introduced. At around Γ point, the bonding and anti-bonding states close to Fermi level are $|s^-\rangle$ and $|p_{xy}^+\rangle$ respectively, with a reversed order [shown in figure 4 (c)]. Without SOC, the system is a zero gap semiconductor. Taking SOC into consideration, the splitted $|p_{xy}^+\rangle$ states lead to an insulating gap as large as 0.3 eV[6]. Given a boundary, the topological nontrivial bulk band would contribute to topological edge states, connecting conduction band and valence band. Figure 4 (d) shows the calculated edge states for stanene with nano-ribbon geometry. And the same helical edges states also exist in fluorinated stanene ribbon (shown in the inset). These results indicate that decorated stanene systems are ideal platforms for large gap 2D TI. Since that, extensive experimental studies are carried out, while many of the experimental results are inconsistent with theoretical prediction. This is because the topological properties of stanene are influenced by many factors, which will be discussed in the next part.

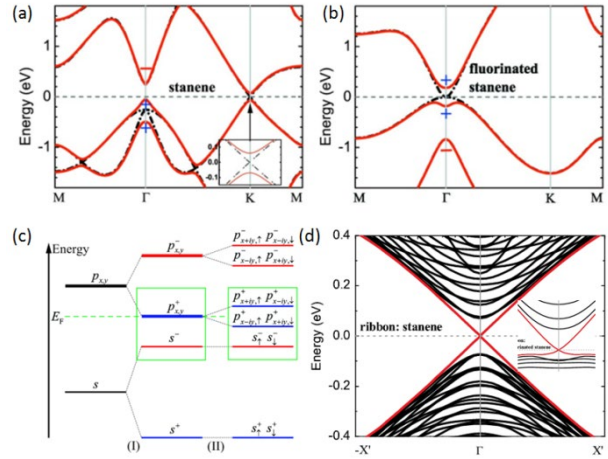

FIG. 4. Topological properties of stanene and decorated stanenes. (a) Band structures of stanene. Band structures at around K point with /without SOC are shown in the inset. Reproduced from Ref [6]. (b) Band structures of decorated stanene. Reproduced from Ref [6]. (c) Schematic diagram of the evolution from the atomic s and $p_{xy}$ orbitals of Sn into the conduction and valence bands at the Γ point for fluorinated stanene. Reproduced from Ref [6]. (d) Edge states for an amchair stanene ribbon. The inset shows the edge states for fluorinated stanene. Reproduced from Ref [6].

### 3.3 Key factors affecting topological properties of stanene

#### 3.3.1 Substrate (stacking manner, strain and lattice constants)

The electronic structure and topological properties of stanene are sensitive to terminated face, lattice strain and chemical composition at the interface with substrate. Plenty of works have been performed to explore the effect of substrates on stanene[7-9, 49-52]. We will give a discussion in the following.

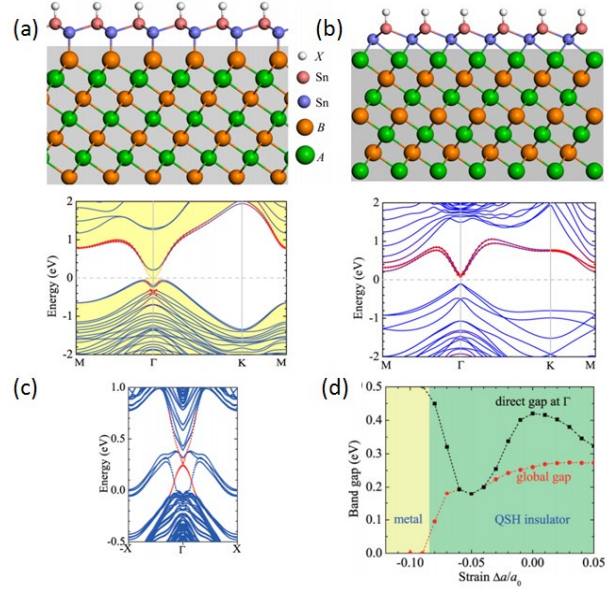

FIG. 5. Influences of substrates on Stanene. (a) Crystal structure of stanene on AB (111)-B type substrate. The band structures are shown at the bottom (taking SrTe as an example). Reproduced from Ref [7]. (b) Crystal structure of stanene on AB (111)-A type substrate. The band structures are shown at the bottom (taking SrTe as an example). Reproduced from Ref [7]. (c) Band structure of a zigzag stanene ribbon grown on AB (111)-B type substrate showing the existence of gapless helical edges (red). Reproduced from Ref [7]. (d) Phase diagram of stanene under strain $\Delta a/a_0$, where $a_0$ is the equilibrium lattice constant of the substrate and $\Delta a$ is the change of the lattice constant. Reproduced from Ref [7].

First of all, the terminated atoms of substrate can extremely affect the geometry configuration of stanene by altering the binding configuration and bonding strength. Taking the AB (111) type substrates (such as SrTe, PbTe, BaSe, and BaTe) for example, the effects of A/B-terminated surface to stanene are discussed. Figure 5 (a) shows the stanene grown on AB (111)-B surface, giving a "top-hcp" stacking configuration. While on AB (111)-A surface, it turns into a "fcc-hcp" configuration [see Fig. 5 (b)]. Band structures for these two configurations are different (shown at each bottom). Stanene in "top-hcp" configuration has an insulating band structure with a band inversion at Γ points, leading to a TI phase with helical edges states (shown in Fig. 5(c)). However, no inverted band structures exist in "fcc-hcp" configuration, giving a trivial insulator phase.

Lattice strain is also an important factor for tuning band structures. For the topological nontrivial case ("top-hcp" configuration), the phase diagram of stanene with lattice strain is shown in Fig. 5 (d). The topological phase of stanene persists under moderate compressive strain or tensile strains, and it changes into semimetallic phase when compressive strain larger than ~8.5 %. In addition, the interlayer distance between stanene film and substrate can also affect electronic properties by modulate the intensity of Rashba splitting[53], which provides another degree of freedom to tune the electronic properties of stanene.

### 3.3.2 Chemical functionalization

Chemical functionalization is also an important method to engineer topological properties of stanene system. It has been studied ever since stanene was born[6, 10]. A lot of chemical functional group, e.g., -H, -F, -Cl, -Br, -I and -OH, can be used to engineer band structures by tuning both lattice constants and $p_z$ orbitals. Table 2 and Figure 6 (a) give some examples of chemical functionalized stanene. And their lattice constants and energy gaps are listed. It is obviously that the lattice is enlarged after chemical decoration, indicating the buckling can be reduced via decoration. Actually, chemical functionalization can enhance the stability of the buckled honeycomb configuration by saturating $p_z$ orbital, leading to a sp3 hybridization for stanene [6]. The band gap enhances from about 0.1 eV to above 0.3 eV. This enhancement is mainly attributed to the gap opening at K points[6], which is mentioned above. Some more complicated chemical functional groups, such as ethynyl-derivatives, can also be used for decoration and give a $SnC_2X$ configuration (X = H, F, Cl, Br, I)[54, 55]. Besides, the functionalized "dumbbell stanene", a transformation from honeycomb structure to dumbbell structure, is calculated and the results show an enlargement of insulating gap[56].

Table 2. Lattice constants and insulating gaps of decorated stanene [10]

| Structure | a (nm) | $E_g$ (eV) |
|---|---|---|
| $SnH_2$ | 0.503 | 0.157 |
| $SnF_2$ | 0.516 | 0.326 |
| $SnCl_2$ | 0.518 | 0.296 |
| SnBr$_2$ | 0.513 | 0.337 |
| SnI$_2$ | 0.515 | 0.343 |

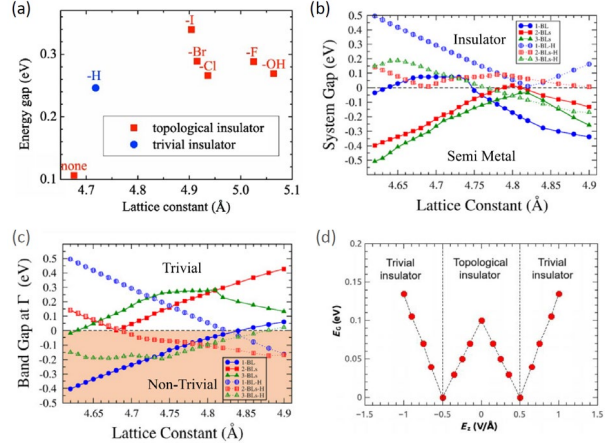

FIG. 6. Influences of different factors to band structures and topological properties of stanene. (a) lattice constants and insulating gaps of different decorated stanene. The topological properties are also marked out. Reproduced from Ref [6]. (b) Calculated global band gaps for different layer stanenes with/ without –H decoration. Reproduced from Ref [11]. (c) Calculated inverted band gaps at Γ point for different layer stanenes with/ without –H decoration. Reproduced from Ref [11]. (d) Electric field induced topological phase transition. Reproduced from Ref [57].

### 3.3.3 Layer dependence

Layer-dependence is general for topological materials, such as WTe$_2$[40], Bi (111)[58] and stanene[11]. The thickness can affect inter-layer coupling, modify the band structure, and even drive a phase transition. Base on the structure of α-Sn, the electronic properties for different layers of stanene are studied. Figure 6 (b) shows the variation of the band gap as a function of lattice constant for monolayer, bilayer and trilayer stanene. In particular, the hydrogenated condition is considered because it is the common case in experiments[12]. The results show that the hydrogenated stanene is more likely to be an insulator. The inverted gap, which determines the topology, is also layer-dependent (see Fig. 6 (c)). According to the results, the monolayer stanene and trilayer hydrogenated-stanene are topological nontrivial in a large range of lattice constant. Considering the generally compressed lattice constant of α-Sn (111) measured in experiments, the trilayer hydrogenated stanene is the optimum selection to realize QSH effect.

### 3.3.4 External electric field

Electric field is effective to tune topological properties for many materials[59-63], including stanene films[57]. According to calculations [see Fig. 6 (d)], external electric field can drive a topological phase transition of stanene films. Without electric field, stanene is a 2D TI. An electric field can decrease the inverted gap. When increasing electric field to $\pm 0.5 \text{ V}/\text{Å}$, the nontrivial gap becomes zero[57]. Further enhancing intensity of the electric field, a trivial gap opens and stanene changes into a trivial insulator[57]. This electric field induced phase transition makes stanene to be a good candidate for controllable topological electronic devices.

### 3.4 Experimental evidences

Although massive theoretical studies and calculation results show that stanene is a good candidate for 2D TI, the sufficient experimental evidences are still lacking. The primary reason is that the structure of stanene is less-stable, compared with other candidates like $WTe_2$, and its electronic properties are sensitive to external influence such as strains and chemical decorations. Thus the strains, induced during the ex-situ fabrication of samples, are nonnegligible and bring disturbances to the measurements. Fortunately, there are a lot of in-situ methods to study stanene films. ARPES can detect the band structure below Fermi level, confirming the nontrivial band structure and insulating gap of stanene. STM has power to resolve the atomically structure of stanene films and give a spatial distribution of edge states. The first experimental band structure of stanene is acquired by using ARPES in 2015, which shows a metallic band without insulating gaps[14]. This is probably due to the compressed lattice constant of stanene grown on $Bi_2Te_3$. Three years later, stanene films in insulating phase is obtained by changing the substrate to PbTe (111) [12] or InSb (111) [13] surface [see Fig. 6 (a) and (b)]. In both experiments, the lattice of stanene is larger than 0.45 nm, which maybe essential for the existence of insulating gap. However, the band inversion is absent in the calculations for both systems. Further progresses are made by using Cu (111) substrates, on which the stanene possesses an inverted band structure and an insulating gap as large as 0.3eV [see Fig. 6 (c)][16]. The STM measurements are also applied to stanene on Cu (111) surface [shown in Fig. 6(d)]. While the edge states are not clearly distinguished in differential conductance (d$I$/d$V$) spectra due to, probably, the influence of Cu substrate.

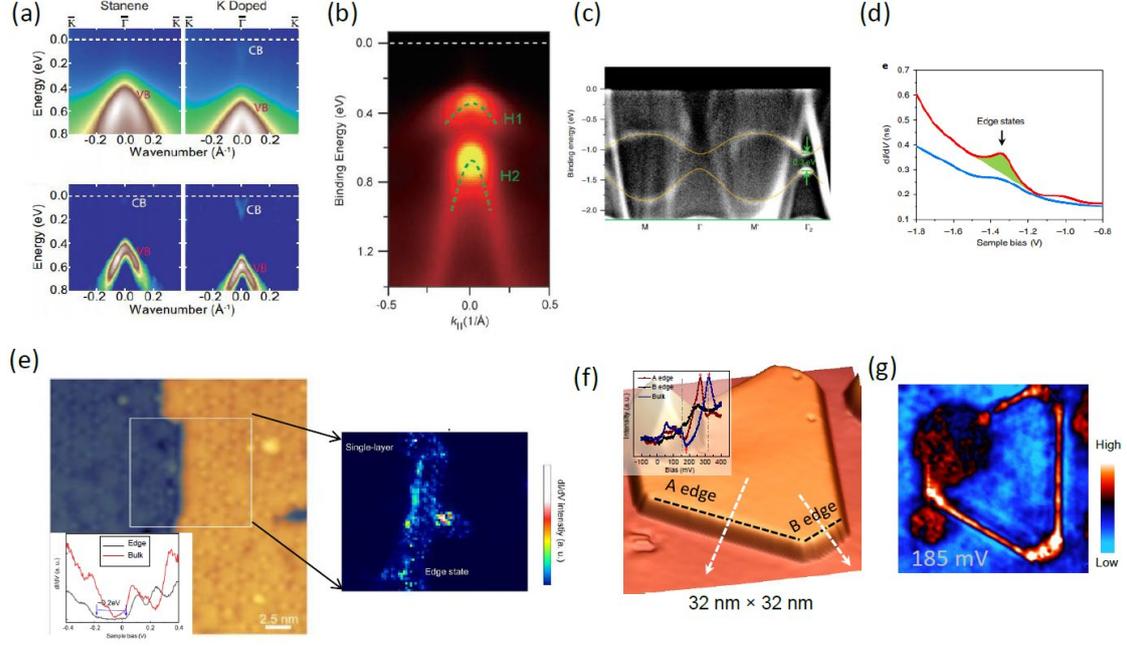

FIG. 7. Experimental results of stanene as a QSH insulator. (a) Band structure of stanene /InSb (111) measured by ARPES before and after K-doping. Reproduced from Ref [13]. (b) Band structure of stanene / PbTe (111) measured by ARPES. Reproduced from Ref [12]. (c) Band structure of both stanene and Cu (111) measured by ARPES, the contribution of stanene is marke out by orange lines. Reproduced from Ref [16]. (d) dI/dV spectra of stanene film [same with the sample in (c)] at center and edge respectively. Reproduced from Ref [16]. (e) STM image of stanene grown on InSb (111). The dI/dV spectra taken in the bulk and at the edges are shown in the inset. The mapping of edge states is shown in the right panel. Reproduced from Ref [17]. (f) STM image of stanene grown on Bi (111). The dI/dV spectra taken at bulk and edges are shown in the inset. Reproduced from Ref [18]. (g) Spatial mapping of edge states (~185 meV) for the same island in (f). Reproduced from Ref [18].

Recently, systematic STM studies of stanene films are reported, by which the spatial distribution of edge states can be observed. The stanene films grown on InSb (111) [17] and Bi (111) [18] substrates are shown in Fig. 7(e) and 7(f), respectively, as well as the dI/d$V$ spectra comparison between bulk and edges. According to the dI/dV spectra one can choose the proper bias for the dI/dV mapping. For stanene/InSb (111) the bias is -100mV, while for stanene/Bi (111) it is 185 mV. This difference is probably que to the different doping effect from the substrates. Both the mapping results show the highly localized edge states at the film edges [see Fig. 7(e) and 7(g)]. These results prove additional evidences for the existence of one dimensional topological edge states in stanene system. However, the smoking gun evidence, the quantized edge conductance $2e^2/h$ in transport measurement, has not yet been observed in stanene system, partially because of the strain-sensitive topological property and partially due to the lacking of suitable insulating substrate to grow stanene.

## 4 Stanene as topological superconductor

In this part, we first give a brief introduction for clarification of TSCs. Then the prediction to realize TSC based on stanene system is discussed. At last, the related experimental evidences reported are introduced.

### 4.1 Classification for topological superconductors

TSC is the superconducting analogue of TI, where the insulating gap is substituted by the superconducting gap. At their boundaries, the self-conjugate excitations called Majorana modes exist, which are blocks for fault-tolerant quantum computation[64]. The most studied type of TSC breaks time reversal symmetry and hosts zero dimensional Majorana bound states at their boundary. This kind of TSC can be realized by inducing superconductivity to ferromagnetic atomic chains or semiconductor nanowires with strong SOC by proximity effect of s-wave superconductors [Fig. 8 (a) and (b)] [65-69]. The second type of TSC is called chiral TSC. Chiral TSC also break time reversal symmetry, but the boundary state changes from localized zero dimensional Majorana bound state to one dimensional chiral Majorana edge mode, which is recently observed in the SC/QAHI heterostructure [(Fig. 8 (c)) [70, 71]. The third type of TSC is proposed to be time reversal invariant, which host Majorana Kramers pairs at boundaries. A simple scenario to realize time-reversal-invariant TSC is to use Ag-doped stanene film[23]. At their boundaries, the helical Majorana modes exist. In addition to these three types of TSC, there is a special type of TSC build by 3D TI/SC heterostructure, which is also time reversal invariant[72]. The topological surface state of TI has p-wave-like pairing by proximity effect[72]. However, there is no such thing as boundaries for this surface. Boundaries appear only when applying an external magnetic field to form vortex, which also breaks time reversal symmetry (see Fig. 8(d))[73, 74]. In the following, we only focus on the 2D time reversal invariant TSC based on stanene system.

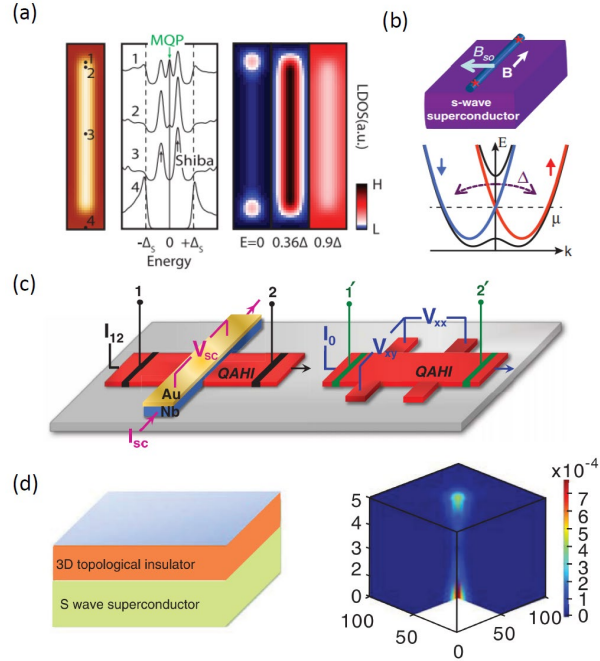

FIG. 8. Different types of TSC. (a) Fe chains on superconductor Pb. The Majorana zero modes are localized at endpoints. Reproduced from Ref [68]. (b) InSb nanowire on a s-wave superconductor. Where the Majorana bound states are localized at endpoints. Reproduced from Ref [67]. (c) Magnetic topological insulator thin film grown on a GaAs (111) B face. Reproduced from Ref [71]. (d) Majorana zero modes in the vortex core of proximity-induced superconducting 3D TI. Reproduced from Ref [74].

## 4.2 Theoretical prediction for stanene to be a 2D topological superconductor

Stanene was first proposed to be a time reversal invariant TSC in 2014 by Zhang *et al* [23]. Base on previous experiments, β-Sn is a superconductor with critical temperature of 3.7 K[75], while the bulk α-Sn shows no superconductivity[24]. It seems that stanene, as a single layer of α-Sn, is not a superconductor. However, the superconductivity of few layer stanene is confirmed by transport measurement[24], which further increases the possibility for stanene to be a TSC.

For decorated stanene, say SnI, the Rashba SOC effect can emerge once breaking the inversion symmetry by, for example, Ag doping [see Fig. 9(a)]. Alternative methods, such as external electric field and heterostructure interface, can also be used to break inversion symmetry. After involving SOC effect, the superconducting pairing symmetry for stanene can be singlet or triplet, depending on the relative intensity of inter-orbital interaction (denoted by *V*) and intro-orbital interaction (denoted by *U*)[23]. The phase diagram of paring symmetry with $U/V$ and $M_0/\mu$ is shown in figure 9 (b), where $M_0$ is the mass parameter concerning band gap of stanene and μ is the Fermi energy. This diagram [Fig. 9 (c)] indicates that triplet pairing is likely to exist when *V* is larger than *U* [23]. And its nontrivial topology is confirmed by Z2 number [Fig. 9(c)]. As a 2D time reversal invariant TSC, there should be helical Majorana edge states residing at the boundaries. Theoretical calculations confirm the existence of this kind of edge states [see Fig. 9(d)].

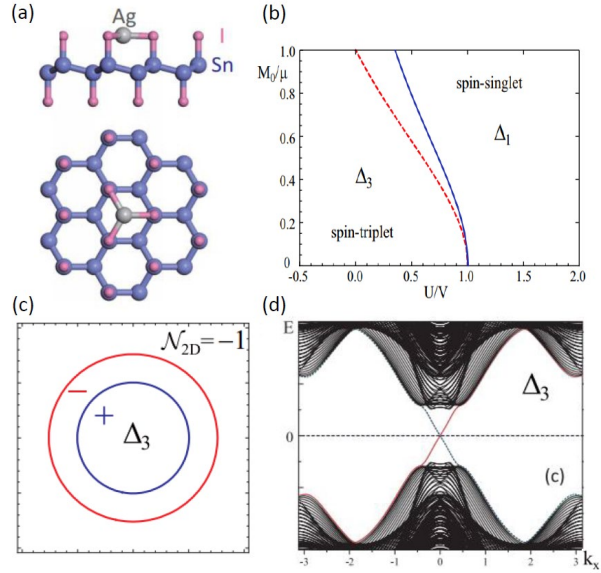

FIG. 9. Triplet pairing in Ag-doped stanene system. (a) Structure of Ag-doped stanene. Side view in upper panel and top view in lower panel. Reproduced from Ref [23]. (b) Phase diagram of pairing symmetry with $U/V$ and $M_0/\mu$. Reproduced from Ref [23]. (c) Bulk Z2 topological number of Ag-doped stanene. Reproduced from Ref [23]. (d) Helical Majorana edge states for triplet pairing condition. Reproduced from Ref [23].

### 4.3 Experimental advancements

Although theoretical predictions show that stanene is a good candidate for time reversal invariant TSC, the experimental evidences are rare. It is mainly because the critical temperature of monolayer stanene is found to be extremely low which requires measurements at ultra-low temperature[18]. Fortunately, this challenge can be solved by increasing the thickness of stanene. Transport measurements show that the critical temperature of superconductivity of stanene depends on film thickness [see Fig. 10(a)] and substrate-induced doping[24]. Thus a suitable substrate and thickness can enhance the critical temperature to a level which is easier to be detected.

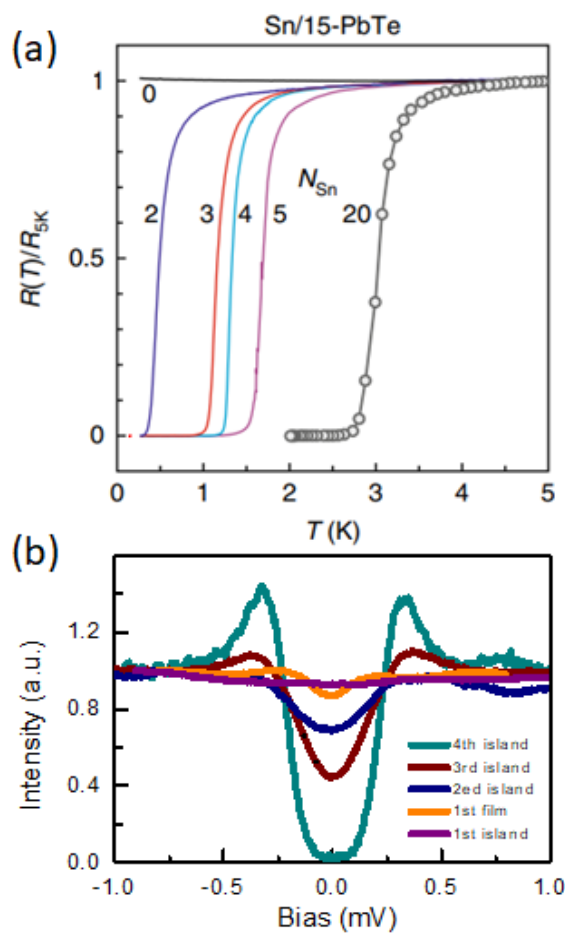

FIG. 10. Experimental results of superconductivity. (a) Transport measurement of superconductivity of few layer stanene/PbTe. Reproduced from Ref [24]. (b) Layer dependence of superconducting gap of stanene. Reproduced from Ref [18].

Recently, the superconductivity of stanene grown on Bi (111) is observed by STM/STS[18]. For monolayer stanene, the superconducting gap is small and shallow even in 350 mk, hindering further investigation for Majorana states. Increasing the thickness of stanene, the superconducting gap gets larger and larger. Four-layer stanene can provide a considerable large superconducting gap, with $\Delta \sim 0.33$ meV [see

Fig. 10 (b)][18], providing a platform to further study triplet pairing and Majorana edge modes.

## 5 Other topological properties in stanene system

Some other topological phases, such as quantum anomalous hall (QAH) insulator and 3D topological Dirac semimetal, can also be realized in stanene system.

### 5.1 QAH insulator based on stanene system

QAH insulator exhibits the same chiral edge mode with quantum hall insulator while without the participation of external magnetic field. Time reversal symmetry in QAH system is broken by spontaneous magnetization[76, 77]. For QAH insulator, the topological order is characterized by Chern number[78]. Chern number determines the number of chiral topological edge states, where electrons can move without backscattering. An advantage for QAH compared with other topological states of matter is that it does not require protection from symmetry. Thus it is an ideal material for dissipationless electronic transport. However, the existing realizations for QAH effect require an extremely low temperature which hinders practical applications[77]. Thus the primary goal in this field is to find materials with large insulating gap and high Curie temperature to realize high-temperature QAH effect.

Magnetic doping into a TI is an effective way to acquire QAH insulator, such as chromium-doped $(Bi,Sb)_2Te_3$[79, 80]. As a large gap 2D TI, stanene is a good candidates for this technique. The Cr-dopoed dumbbell stanene is proposed to a QAH insulator [see Fig.11(c)] [81]. What is more, the gap size for this system is tunable by external strain, which can be as large as 50 meV [81]. While for this scheme, disorders caused by doped atoms can destroy the predicted phenomenon in theoretical prediction. An alternative scheme is to use chemical functionalization to overcome this drawback. The half-passivated stanene is possible to evolve into the ferromagnetic order phase, and the system become a QAH insulator [see Fig. 11 (a) and (b)] [82]. Although this scheme avoids any magnetic doping or magnetic substrate, it is hard to precisely control the position of passivated adatoms. Thus the realization of QAH effect in stanene based system has not realized yet.

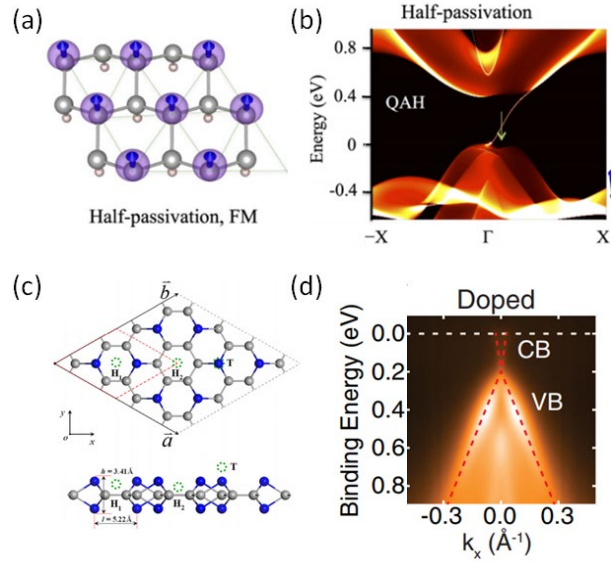

FIG. 11. Schemes to realize QAH insulator and 3D topological Dirac semimetal base on stanene system. (a) Achieving ferromagnetic order in stanene system via functionalization. Reproduced from Ref [82]. (b) Band structure for stanene in QAH phase. Reproduced from Ref [82]. (c) Cr-doped stanene. Reproduced from Ref [81]. (d) Band dispersion of α-Sn/InSb after electron doping. Reproduced from Ref [83].

### 5.2 Realizing 3D Dirac semimetal

Besides QAH insulator, multilayer stanene provides a platform for 3D topological Dirac semimetal, such as multilayer stanene grown on InSb[83]. Single layer stanene on InSb substrate has a large insulating gap ~ 0.44eV[13]. Increasing the thickness up to 6 layers, it becomes Dirac semimetal with a pair of Dirac cones exist along $k_z$ axis near Γ point. ARPES data confirms these Dirac cones in potassium doped 6-BL α-Sn film/InSb sample [see Fig. 11 (d)][83]. What is more, a phase transition from TI to Dirac semimetal can be tuned by a slight in-plane strain (less than 1%), which offers great potential for device applications.

## 6 Summary and perspective

As a topological material with plenty of exotic properties, stanene has been intensively studied, mainly as a candidate for QSH insulator and TSC. Massive efforts in both theory and experiment have been made. However, for all kinds of difficulties, the conclusive evidences are still lacking. As a QSH insulator, stanene has shown the inverted band gap on Cu (111) substrate and the edge states strictly residing at the film edges on InSb(111) and Bi (111) substrates. However, these systems have their disadvantages: both of Cu and Bi substrates are metallic; too many defects are found in stanene films grown on InSb. All these hinder the further transport measurements of the quantized edge conducting channels. Finding a suitable insulating substrate to grow stanene is the problem to be solved immediately. As a TSC, the superconductivity of few layers stanene is confirmed by transport measurements and STM, respectively. But its topological properties are still unknown, which needs to be addressed by further experiments. Moreover, signatures of Majorana edge modes are absent. The dispersing edge states similar with that recently reported residing at domain wall of $FeSe_{0.45}Te_{0.55}$ [84] are expected to be observed.

In general, the nontrivial topological properties, large gap character as well as the abundant tuning methods make stanene a fascinating material for fundamental researches and practical applications.

**Acknowledgement**

We acknowledge the financial support from National Natural Science Foundation of China (Grant Nos. 11521404, 11634009, 11674222, 11674226, 11790313, 11574202, 11874256, U1632102, 11861161003 and 11874258), the National Key Research and Development Program of China (Grant Nos. 2016YFA0300403, 2016YFA0301003). This work is also supported in part by the Key Research Program of the Chinese Academy of Sciences (Grant No. XDPB08-2), the Strategic Priority Research Program of Chinese Academy of Sciences (Grant No. XDB28000000).